\documentclass[12pt]{article}
\usepackage{amssymb}
\usepackage{amsmath}

\input epsf
\input{tcilatex}
\begin{document}

\title{\textbf{Fsusy and Field Theoretical Construction}}
\author{}
\maketitle

\begin{center}
\bigskip \textbf{M.B. SEDRA},\footnote{%
Corresponding author: msedra@ictp.it} and \textbf{J. ZEROUAOUI} \\[0pt]
%EndAName
{\small Universit\'{e} Ibn Tofail, Facult\'{e} des Sciences, D\'{e}partement
de Physique,}

{\small Laboratoire de Physique de la Mati\`{e}re et Rayonnement (LPMR), K%
\'{e}nitra, Morocco}\\[0pt]
\end{center}
\begin{abstract}
Following our previous work on fractional spin symmetries (FSS)
\cite{sz1,sz2}, we consider here the construction of field
theoretical models that are invariant under the $D=2(1/3,1/3)$
supersymmetric algebra.
\end{abstract}

\section{Superspace setup}

Fractional supersymmetry \cite{1, ssz1, ssz2, 4, 5, sz1, sz2} is
once again considered in this work. The $D=2(1/3,1/3)$
superalgebra discussed in \cite{ssz1, ssz2, sz1, sz2} is generated
by the left and right conserved charges $Q_{1/3}^{-}$, $Q_{1/3}^{+}$, $P$ and $%
\overline{Q}_{-1/3}^{-}$, $\overline{Q}_{-1/3}^{+}$, $\overline{P}$ \
respectively together with four topological charges $\Delta ^{(-,-)},$ $%
\Delta ^{(-,+)},$ $\Delta ^{(+,-)}$ and $\Delta ^{(+,+)}$ relating the two
sectors. The $\pm $ and $0$ charges carried by these objects are those of
the $Z_{3}\times \overline{Z}_{3}$ automorphism symmetry acting as:
\begin{eqnarray}
\Gamma Q^{+} &=&{q}Q^{+},\text{ }\Gamma Q^{-}=\overline{q}Q^{-},\text{ }%
\Gamma P=P,  \notag \\
\overline{\Gamma }\overline{Q}^{+} &=&{q}\overline{Q}^{+},\text{ }\overline{%
\Gamma }\overline{Q}^{-}=\overline{q}\overline{Q}^{-},\text{ }\overline{%
\Gamma }\overline{P}=\overline{P}, \\
\overline{\Gamma }\overline{Q}^{\pm } &=&\overline{Q}^{\pm },\text{ }%
\overline{\Gamma }Q^{\pm }=Q^{\pm },\text{ }\Gamma \overline{P}=\overline{P},
\notag
\end{eqnarray}%
where $\Gamma $ and $\overline{\Gamma }$ are the generators of the $Z_{3}$
and $\overline{Z}_{3}$ group and where we have used the convention notations
$Q_{-1/3}^{\pm }=\overline{Q}^{\pm }$ and $P_{-1}=\overline{P}$ in addition
to $Q_{1/3}^{\pm }=Q^{\pm }$ and $P_{1}=P$ used in \cite{sz2}. The $%
D=2(1/3,1/3)$ supersymmetric algebra admits moreover an extra $Z_{2}\times
\overline{Z}_{2}$ symmetry, generated by $C\otimes \overline{C}$, acting as
follows:
\begin{eqnarray}
CQ^{-} &=&Q^{+}C,  \notag \\
\overline{C}\overline{Q}^{-} &=&\overline{Q}^{+}\overline{C}.
\end{eqnarray}%
Note that under the complex conjugation ($\ast $) of complex variables $%
z^{\ast }=\overline{z},$ we have the obvious relations:
\begin{eqnarray}
\overline{Q}^{+} &=&\left( Q^{-}\right) ,  \notag \\
\overline{P} &=&P^{\ast },
\end{eqnarray}%
showing that the left and right sectors are related by the complex
conjugation of the two dimensional world sheet parametrized by $z$ and $%
\overline{z}$. For later use, we quote hereafter the different
automorphisms of the $D=2(1/3,1/3)$ superalgebra
\begin{equation}
\begin{tabular}{|c|c|c|c|c|c|c|c|}
\hline
& $Q^{+}$ & $Q^{-}$ & $\overline{Q}^{+}$ & $\overline{Q}^{-}$ & $P$ & $%
\overline{P}$ & $\Delta ^{(-,-)}$ \\ \hline
$C$ & $Q^{-}$ & $Q^{+}$ & $\overline{Q}^{+}$ & $\overline{Q}^{-}$ & $P$ & $%
\overline{P}$ & $\Delta ^{(+,-)}$ \\
$\overline{C}$ & $Q^{+}$ & $\overline{Q}^{-}$ & $\overline{Q}^{-}$ & $%
\overline{Q}^{+}$ & $P$ & $\overline{P}$ & $\Delta ^{(-,+)}$ \\
$\ast $ & $\overline{Q}^{-}$ & $\overline{Q}^{+}$ & $Q^{-}$ & $Q^{+}$ & $%
\overline{P}$ & $P$ & $\Delta ^{(+,+)}$ \\ \hline
\end{tabular}%
\end{equation}%
A differential representation of the $D=2(1/3,1/3)$
superalgebra respecting (4) may be obtained by introducing a large superspace $(z,$ $%
\theta ^{\pm },$ $\overline{z},$ $\overline{\theta }^{\pm },$ $x^{++},$ $%
x^{--},$ $x^{+-},$ $x^{-+})$ with $\left( \theta ^{\pm }\right) ^{3}=0$ and $%
\left( \overline{\theta }^{\pm }\right) ^{3}=0.$ We find
\begin{eqnarray}
\emph{D}^{-} &=&D^{-}+\alpha \overline{\theta }^{-}\partial
^{(-,+)}+\alpha
\overline{\theta }^{+}\partial ^{(-,-)},  \notag \\
\emph{D}^{+} &=&D^{+}+\alpha \overline{\theta }^{+}\partial
^{(+,-)}+\alpha
\overline{\theta }^{-}\partial ^{(+,+)},  \notag \\
\overline{\emph{D}}^{-} &=&\overline{D}^{-}+\overline{\alpha
}\theta ^{-}\partial
^{(+,-)}+\overline{\alpha }\theta ^{+}\partial ^{(-,-)},  \notag \\
\overline{\emph{D}}^{+} &=&\overline{D}^{+}+\overline{\alpha
}\theta ^{+}\partial
^{(-,+)}+\overline{\alpha }\theta ^{-}\partial ^{(+,+)},   \\
P &=&-\overline{q}\frac{\partial }{\partial z},\overline{P}=-q\frac{\partial
}{\partial \overline{z}},  \notag \\
\Delta ^{(+,+)} &=&(\overline{\alpha }-\alpha q)\partial ^{(+,+)},\Delta
^{(-,-)}=(\overline{\alpha }-\alpha \overline{q})\partial ^{(-,-)},  \notag
\\
\Delta ^{(+,-)} &=&(\overline{\alpha }-\alpha \overline{q})\partial
^{(+,-)},\Delta ^{(-,+)}=(\overline{\alpha }-\alpha q)\partial ^{(-,+)},
\notag
\end{eqnarray}%
where the derivatives along the extra directions, realizing the topological charges as translation
generators, are defined as: $\partial ^{(+,+)}=\frac{\partial }{\partial
x^{--}}$\ and so on. $\overline{D}^{-}$ and $\overline{D}^{+}$ are the spin $%
\frac{1}{3}$ charge operators realizing the right sector of the
$D=2(1/3,1/3)$ superalgebra without topological charges. As shown
on the table (4), $\overline{D}^{-}$ and $\overline{D}^{+}$ read
as :
\begin{eqnarray}
\overline{D}^{+}&=&\partial/\partial {\overline
{\theta}^-}+{\overline{\theta}^{-2}}\partial/\partial\bar z \notag \\
\overline{D}^{-}&=&\partial/\partial
\bar{\theta^+}+\overline{\theta}^{+2}\partial/\partial\bar z
\end{eqnarray}
Note that $\overline{D}^{-}$ and $\overline{D}^{+}$ are related to
each other the $\overline{Z}_{2}$  automorphism group acting on
the superspace variables $\theta^\pm$ and $z$ as
\begin{eqnarray}
\overline{C}\overline{\theta }^{-} &=&\overline{\theta }^{+}\overline{C} \notag \\
\overline{C}\overline{z} &=&\overline{z}\overline{C}
\end{eqnarray}%
Superfields describing off shell representations of the
$D=2(1/3,1/3)$ superalgebra are
superfunctions defined on the generalized superspace $(z,\theta ,\overline{z}%
,\overline{\theta },x)$. They consist of $3^{4}=81$ component
fields depending on the bosonic variable $z,$ $\overline{z}$ and
$x$. This is a big number of degrees of freedom that renders very
difficult the elaboration of invariant field theoretical models
under the $D=2(1/3,1/3)$ symmetry. However, forgetting about some
automorphism symmetries, one may construct models which are
invariant under subalgebras of $D=2(1/3,1/3)$. To do that, various
possibilities are in order. The simple way is to ignore all the
automorphisms given by (4). This is the case of the subalgebra:
\begin{eqnarray}
Q^{-3} &=&P,  \notag \\
\overline{Q}^{-3} &=&\overline{P}, \\
Q^{-}\overline{Q}^{-}-\overline{q}\overline{Q}^{-}Q^{-} &=&\Delta ^{(-,-)}
\notag
\end{eqnarray}%
generated by non hermitian charge operators. Non unitary invariant
models under this symmetry will be considered here. The same thing
may be
said about the subalgebra generated by $\left( Q^{+},\overline{Q}^{+},P,%
\overline{P}\right) $ as it is related to (8) by the $Z_{2}\times \overline{Z%
}_{2}$ symmetry generated by $C\otimes \overline{C}.$ The second
kind of models, which will be studied later, are based on the
subalgebra:
\begin{eqnarray}
\overline{Q}^{-} &=&P,  \notag \\
\overline{Q}^{+} &=&\overline{P}, \\
Q^{-}\overline{Q}^{+}-q\overline{Q}^{+}Q^{-} &=&\Delta .  \notag
\end{eqnarray}
These equations are stable under the complex ($\ast $) as shown by
(4). The field theory invariant under (9) is real and may describe
unitary $D=2(1/3,1/3)$ supersymmetric models. In what follows, we
first study those theories that are invariant under (8).
Introducing the superspace $(z,\theta
^{+},\overline{z},\overline{\theta }^{+},x^{++})$ with $\theta ^{+3}=0$ and $%
\overline{\theta }^{+3}=0$ a representation of this algebra reads as:
\begin{eqnarray}
D^{-} &=&D^{-}+\alpha \overline{\theta }^{+}\partial ^{(-,-)},  \notag \\
\overline{D}^{-} &=&\overline{D}^{-}+\beta \theta ^{+}\partial ^{(-,-)},
\notag \\
P &=&-\overline{q}\frac{\partial }{\partial z},\overline{P}=-q\frac{\partial
}{\partial \overline{z}}, \\
\Delta ^{(-,-)} &=&(\beta -\alpha \overline{q})\partial ^{(-,-)},
\notag
\end{eqnarray}
where $D^{-}$ and $\overline{D}^{-}$ are given by (6). To check
that the above relations form indeed a representation of (8), we
follow the same strategy as before. First we calculate the the square of $D^{-}$($\overline{D%
}^{-}$). We find:%
\begin{equation}
D^{-2}=D^{-2}+\overline{\theta }^{+2}\partial ^{(-,-)2}+(1+\overline{q})%
\overline{\theta }^{+}\partial ^{(-,-)}D^{-},
\end{equation}%
where $D^{-2}$ is given by \cite{sz2}
\begin{equation}
{D{^{-2}=\partial }}^{2}{/{\partial \theta ^{+2}+(1+q)\theta ^{+}\partial }/{%
\partial \theta ^{+}\partial }/{\partial z+(1+q}}^{2}){{\theta ^{+2}\partial
}}^{2}{/{\partial \theta ^{+2}\partial }/{\partial z}}
\end{equation}
Repeating the some procedure, we get:
\begin{equation}
D^{-3}=(1+q)\frac{\partial }{\partial z}+\alpha (1+\overline{q}+\overline{q}%
^{2})[\overline{\theta }^{+}D^{-}+\alpha \overline{\theta }^{+2}\partial
^{(-,-)}]D^{-}\partial ^{(-,-)},
\end{equation}%
which reduces to $P$ because of the identity $1+\overline{q}+\overline{q}%
^{2}=0$. A similar proof is valid for $\overline{D}^{-}$. Moreover using the
commutation rules:
\begin{eqnarray}
D^{-}\overline{D}^{-} &=&q\overline{D}^{-}D^{-} \notag\\
D^{-}\overline{\theta }^{+} &=&\overline{q}\overline{\theta }^{+}D^{-}, \notag\\
\overline{D}^{-}\theta ^{+} &=&\overline{q}\overline{\theta }^{+}\overline{D}%
^{-} \\
P\overline{D}^{-} &=&\overline{D}^{-}P \notag\\
\Delta ^{(-,-)}\overline{D}^{-} &=&\overline{D}^{-}\Delta
^{(-,-)}, \notag
\end{eqnarray}%
It is not difficult to see that the second equality of (8) is also
satisfied.
\section{Field theoretical construction}
Superfields defined on the superspace $(z,\theta ^{+},\overline{z}%
,\overline{\theta }^{+},x^{++})$ are usually complex. These off shell
representations, consisting of $3^{2}$ complex degrees of freedom, may carry
both a spin $s=h-\overline{h}$ and $Z_{3}\times \overline{Z}_{3}$ charge $%
(m,n)$ with $m,n=0,\pm 1({mod}3).$ The $\theta _{-1/3}^{+}$ and $\overline{%
\theta }_{1/3}^{+}$ expansion of a generic superfield $\phi _{h,\overline{h}%
}^{(m,n)}$ reads as:
\begin{eqnarray}
\phi _{h,\overline{h}}^{(m,n)} &=&\varphi _{h,\overline{h}}^{(m,n)}+\theta
_{-1/3}^{+}\psi _{(h+\frac{1}{3},\overline{h})}^{(m-1,n)}+\overline{\theta }%
_{1/3}^{+}\eta _{(h,\overline{h}+\frac{1}{3})}^{(m,n-1)} \notag
\\
&+&\theta _{-1/3}^{+}\overline{\theta }_{1/3}^{+}F_{(h+\frac{1}{3},\overline{%
h}+\frac{1}{3})}^{(m-1,n-1)}+\theta _{-1/3}^{+2}\chi _{(h+\frac{2}{3},%
\overline{h})}^{(m-2,n)}  \notag \\
&+&\overline{\theta }_{1/3}^{+2}\lambda _{(h,\overline{h}+\frac{2}{3}%
)}^{(m,n-2)}+\theta _{-1/3}^{+2}\overline{\theta }_{1/3}^{+}\xi _{(h+\frac{2%
}{3},\overline{h}+\frac{1}{3})}^{(m-2,n-1)}  \notag \\
&+&\overline{\theta }_{1/3}^{+2}\theta _{-1/3}^{+}V_{(h+\frac{1}{3},%
\overline{h}+\frac{2}{3})}^{(m-1,n-2)}+\theta _{-1/3}^{+2}\overline{\theta }%
_{1/3}^{+2}D_{(h+\frac{2}{3},\overline{h}+\frac{2}{3})}^{(m-2,n-2)}
\end{eqnarray}%
Taking $h=\overline{h}=-1$ and $n=m=-1$, we that the fields
$\varphi
^{(-,-)} $, $\chi _{\frac{2}{3}}^{(0,-)}$, $\lambda _{-\frac{2}{3}}^{(-,0)}$%
and $D^{(-,-)}$ are exactly those appearing in the realization of thee
critical spin $\frac{1}{3}$ supersymmetry of the TPM namely
\begin{eqnarray}
\varphi ^{(-,-)} &=&\phi _{\frac{1}{21},\frac{1}{21}}^{(-,-)}~\
,D^{(0,0)}=\phi _{\frac{1}{21},\frac{5}{7}}^{(0,0)}\ \notag \\
\chi _{\frac{2}{3}}^{(0,-)} &=&\phi
_{\frac{5}{7},\frac{1}{21}}^{(0,-)}~\ ,\lambda
_{\frac{-2}{3}}^{(0,-)}=\phi _{\frac{1}{21},\frac{5}{7}}^{(-,0)}~\
\end{eqnarray}%
The remaining fields $\psi _{\frac{1}{3}}^{(-,-)}$, $\eta _{-\frac{1}{3}%
}^{(-,+)}$, $F^{(+,+)},$ $\xi _{\frac{1}{3}}^{(0,+)}$ and $V_{-\frac{1}{3}%
}^{(+,0)}$ which are identified with:
\begin{eqnarray}
\psi _{\frac{1}{3}}^{(-,-)} &=&\phi _{\frac{8}{21},\frac{1}{21}}^{(+,-)}~\
,\eta _{-\frac{1}{3}}^{(-,+)}=\phi _{\frac{1}{21},\frac{8}{21}}^{(-,+)}\  \notag\\
\xi _{\frac{1}{3}}^{(0,+)} &=&\phi
_{\frac{5}{7},\frac{8}{21}}^{(0,+)}~\
,F^{(+,+)}=\phi _{\frac{8}{21},\frac{8}{21}}^{(+,+)} \\
V_{-\frac{1}{3}}^{(+,0)} &=&\phi
_{\frac{8}{21},\frac{5}{7}}^{(+,0)}\notag
\end{eqnarray}%
Are extra conformal fields since they are not predicted by the
$C=\frac{6}{7} $ conformal theory. They are however indispensable
in the building of a manifestly $D=2(1/3,1/3)$ supersymmetric
theory eventually invariant under the $Z_{3}\times
\overline{Z}_{3}$ discrete symmetry of (8). As for the left
sector considered previously, here also the highest $\theta
$-component terms of superfields (14) transform as a total space
time derivative under
the change $\delta \theta ^{+}=\varepsilon ^{+}$ and $\delta \overline{%
\theta }^{+}=\overline{\varepsilon }^{+}$. Invariant actions $S$ are then
constructed as in $D=2(1/2,1/2)$ supersymmetric theories. We have:\bigskip
\begin{equation}
S=\int d^{2}zd^{4}\theta L^{(-,-)},
\end{equation}%
where $\int d^{4}\theta \sim \overline{D}^{-2}D^{-2}$ and where the
super-Lagrangian $L^{(-,-)}$ carries a $(-,-)Z_{3}\times \overline{Z}_{3}$
charge and scales as $1/3+1/3$ dimensional quantity since the integral
measure scales as $($length$)^{1/3+1/3}$. Note that we have ignored the $x-$%
dependence realizing the topological charge $\Delta ^{(-,-)}$. Other details
will given when examining hermitian models. Using dimensional arguments, it
is not difficult to see that $L^{(-,-)}$ is of the form: \bigskip
\begin{equation}
L^{(-,-)}\sim D^{-}\phi _{1}^{(m,n)}\overline{D}^{-}\phi
_{2}^{(-m,-n)}+W^{(-,-)}(\phi _{1},\phi _{2}),
\end{equation}%
where the integers $m,n$ may take the values $0,\pm 1$. As pointed out from
the beginning of this work, the action $S$ and the lagrangian (18-19) are
not hermitian. Setting $n=m=-1$ as suggested by the thermal deformation of
the TPM \cite{sz1} and using the $\theta $-expansion of the complex
superfields $\phi _{1}^{(-,-)}$ and $\phi _{2}^{(+,+)}$ as well as the
expression of the derivatives $D^{-}$ and $\overline{D}^{-}$ and
(6), one may calculate the component fields contribution to the action $S$
of the first term of (18). Straightforward algebra leads to:
\bigskip
\begin{eqnarray}
D^{-}\phi _{1}^{(-,-)} &=&\psi _{\frac{1}{3}}^{(+,-)}+\overline{\theta }%
^{+}F^{(+,+)}+q^{2}\overline{\theta }^{+2}V_{-\frac{1}{3}}^{(+,0)}  \notag \\
&&-\overline{q}\theta ^{+}\left( \chi _{\frac{2}{3}}^{(0,+)}+\overline{%
\theta }^{+}\xi _{\frac{1}{3}}^{(0,+)}+\overline{\theta }^{+2}D^{(0,0)}%
\right)  \notag \\
&&+\theta ^{+2}\left( \partial \varphi +\overline{\theta }^{+}\partial \eta
_{-\frac{1}{3}}^{(-,+)}+\overline{\theta }^{+2}\partial \lambda _{-\frac{2}{3%
}}^{(-,0)}\right) ,
\end{eqnarray}
\begin{eqnarray}
\overline{D}^{-}\phi _{1}^{(+,+)} &=&\overline{\eta }_{-\frac{1}{3}}^{(+,0)}+%
\overline{q}\theta ^{+}\overline{F}^{(0,0)}+q\theta ^{+2}\overline{\xi }_{%
\frac{1}{3}}^{(-,0)}  \notag \\
&&-q\overline{\theta }^{+}\left( \overline{\lambda }_{-\frac{2}{3}%
}^{(+,-)}+\theta ^{+}\overline{V}_{-\frac{1}{3}}^{(0,-)}+q\theta ^{+2}%
\overline{D}^{(-,-)}\right)  \notag \\
&&+\overline{\theta }^{+2}\left( \overline{\partial }\overline{\varphi }%
^{(+,+)}+\theta ^{+}\overline{\partial }\overline{\psi }_{\frac{1}{3}%
}^{(0,-)}+\theta ^{+2}\overline{\partial }\overline{\chi }_{\frac{2}{3}%
}^{(-,+)}\right) ,
\end{eqnarray}
where we have put a bar on the component fields of the superfield $\phi
_{2}^{(+,+)}$ in order to avoid the confusion with the $\phi _{1}^{(-,-)}$
component fields. Evidently, $\partial $ and $\overline{\partial }$ mean $%
\frac{\partial }{\partial z}$ and $\frac{\partial }{\partial \overline{z}}$
respectively. Integrating with respect to $d^{4}\theta $ the superfield
kinetic term taking into account the commutation rule $\overline{\theta }%
^{+}\theta ^{+}=q\theta ^{+}\overline{\theta }^{+}$, we get:
\begin{eqnarray}
L_{0} &\sim &\partial \varphi ^{(-,-)}\overline{\partial }\overline{\varphi }%
^{(+,+)}+(\partial \lambda _{-2/3}^{(-,0)}\overline{\eta }_{-1/3}^{(+,0)}-%
\overline{q}\chi _{2/3}^{(0,-)}\overline{\partial }\overline{\psi }%
_{1/3}^{(0,-)}) \notag\\
&&\text{ }-\left( \partial \eta _{-\frac{1}{3}}^{(-,+)}\overline{\lambda }_{-%
\frac{2}{3}}^{(+,-)}-\overline{q}\psi _{1/3}^{(+,-)}\overline{\partial }%
\overline{\chi }_{\frac{2}{3}}^{(-,+)}\right) \notag\\
&&-\overline{q}\psi -qD^{(0,0)}\overline{F}^{(0,0)}-\overline{q}F^{(+,+)}%
\overline{D}^{(-,-)}+\overline{q}V_{-1/3}^{(+,0)}\overline{\xi }%
_{1/3}^{(-,0)}+\overline{q}\xi _{1/3}^{(0,+)}\overline{V}_{-1/3}^{(0,-)}.
\end{eqnarray}%
Note that this relation contains two kinds of fields. Dynamical fields
namely $\varphi ^{(-,-)},\psi ^{(+,-)},$ $\overline{\eta }^{(-,+)},$ $%
\lambda ^{(-,0)}$ and $\overline{\varphi }^{(+,+)},$ $\overline{\psi }%
^{(0,-)},$ $\overline{\overline{\eta }}^{(+,0)},$ $\overline{\lambda }%
^{(+,0)}$. The obey free field equations of motion whose solutions
factorise
into analytic and antianalytic parts. Auxiliary fields $F^{(+,+)},$ $%
V^{(+,0)},$ $\xi ^{(0,+)},$ $D^{(0,0)}$ and $\overline{F}^{(0,0)},$ $%
\overline{\xi }^{(-,0)},$ $\overline{V}^{(0,-)},$
$\overline{D}^{(-,-)}$. They appear linearly in $L_{0}$ and lead
then to constraint equations. Details on the role of these fields
will be given later. Note moreover that (22) is invariant under the
following transformations:
\begin{eqnarray}
\delta \varphi ^{(m,n)} &=&\varepsilon _{-1/3}^{+}\psi _{1/3}^{(m-1,n)}+%
\overline{\varepsilon }_{1/3}^{+}\eta _{-1/3}^{(m,n-1)}  \notag \\
\delta \psi _{1/3}^{(m-1,n)} &=&-q\varepsilon _{-1/3}^{+}\chi
_{2/3}^{(m-2,n)}+\overline{\varepsilon }_{1/3}^{+}F^{(m-1,n-1)}  \notag \\
\delta \eta _{-1/3}^{(m,n-1)} &=&-\overline{\varepsilon }_{1/3}^{+}\lambda
_{-2/3}^{(m,n-2)}+\overline{q}\varepsilon _{-1/3}^{+}F^{(m-1,n-1)}  \notag \\
\delta F^{(m-2,n-1)} &=&(1+q)\varepsilon _{-1/3}^{+}\xi _{1/3}^{(m-2,n-1)}-%
\overline{q}\overline{\varepsilon }_{1/3}^{+}V_{-1/3}^{(m-1,n-2)}  \notag \\
\delta \chi _{2/3}^{(m-2,n)} &=&q\varepsilon _{-1/3}^{+}\partial \varphi
^{(m,n)}+\overline{\varepsilon }_{1/3}^{+}\xi _{1/3}^{(m-2,n-1)} \\
\delta \lambda _{-2/3}^{(m,n-2)} &=&\overline{q}\overline{\varepsilon }%
_{1/3}^{+}\overline{\partial }\varphi ^{(m,n)}+\varepsilon
_{-1/3}^{+}V_{-1/3}^{(m-1,n-2)}  \notag \\
\delta \xi _{1/3}^{(m-2,n-1)} &=&\overline{q}\varepsilon _{-1/3}^{+}\partial
\eta _{-1/3}^{(m,n-1)}-q\overline{\varepsilon }_{1/3}^{+}D^{(m-2,n-2)}
\notag \\
\delta V_{-1/3}^{(m-1,n-2)} &=&\overline{q}\overline{\varepsilon }_{1/3}^{+}%
\overline{\partial }\psi _{1/3}^{(m-1,n)}-q\varepsilon
_{-1/3}^{+}D^{(m-2,n-2)}  \notag \\
\delta D^{(m-2,n-2)} &=&\varepsilon _{-1/3}^{+}\partial \lambda
_{-2/3}^{(m,n-2)}-\overline{q}\overline{\varepsilon }_{1/3}^{+}\overline{%
\partial }\chi _{2/3}^{(m-2,n)}  \notag
\end{eqnarray}%
The spin $\pm 4/3$ supersymmetric conserved current $G^{-}$and $\overline{G}%
^{-}$ generating these transformations are obtained by using the Noether
method. They read as:%
\begin{eqnarray}
G^{-} &=&\partial \varphi ^{(-,-)}\overline{\psi }^{(0,+)}-q\psi
^{(+,0)}\partial \overline{\varphi }^{(+,+)}+\overline{q}\chi ^{(0,-)}%
\overline{\chi }^{(-,+)}, \\
\overline{G}^{-} &=&\overline{q}\overline{\partial }\varphi ^{(-,-)}%
\overline{\eta }^{(+,0)}+\overline{q}\overline{\partial }\overline{\varphi }%
^{(+,+)}\eta ^{(-,+)}+\lambda ^{(-,0)}\overline{\lambda }^{(+,-)}
\end{eqnarray}
Finally, observe that starting from (23, 24) and using the $Z_{2}$%
-symmetries generated by $C$ and $\overline{C}$, we can build the field
realisations of the dual current $G^{+}$ and $\overline{G}^{+}$ as follows:%
\begin{equation}
G^{+}=CG^{-}C^{-1},\overline{G}^{+}=\overline{C}\overline{G}^{-}\overline{C}%
^{-1}.
\end{equation}%
\bigskip We have for $G^{+}$ for instance:
\begin{equation}
G^{+}=\partial \varphi ^{(+,-)}\psi ^{(0,+)}+\psi _{1/3}^{(-,-)}\partial
\overline{\varphi }^{(-,+)}+\overline{q}\chi _{2/3}^{(0,-)}\overline{\chi }%
_{2/3}^{(+,+)}+...,
\end{equation}%
where $C\varphi ^{(+,-)}=\varphi ^{(-,-)}C$ and so on. The superpotentiel term $%
W^{(-,-)}$ is a priori an arbitrary function of the superfields $\phi _{1}$
and $\phi _{2}$ which may be restricted by requiring convariance under the $%
Z_{3}\times \overline{Z}_{3}$ transformations. The most general form of $%
W^{(-,-)}$ respecting the $Z_{3}\times \overline{Z}_{3}$ symmetry reads then
as%
\begin{eqnarray}
W^{(-,-)} &=&\sum\limits_{m}\left( g_{m}\phi
_{1}^{(-,-)^{3m+1}}+g_{m}^{^{\prime }}\phi _{2}^{(+,+)^{3m+2}}\right)  \notag
\\
&&+\sum\limits_{m}g_{m}^{^{\prime \prime }}\phi _{1}^{(-,-)^{m+2}}\phi
_{2}^{(+,+)^{m+1}}
\end{eqnarray}%
where $g_{m}$, $g_{m}^{\prime }$ and $g_{m}^{^{\prime \prime }}$ are
coupling constants. Note that leading linear term in the above equation $%
g_{0}\phi _{1}^{(-,-)}$, integrated with respect to $d^{4}\theta ,$ give $%
g_{0}D^{(0,0)}.$ It describes exactly the $\phi _{1,3}$ thermal
perturbation of the $c=6/7$ critical theory as shown by (16) and
(14). We expect that this term is the mediator of the spontaneous
breaking of the $D=2(1/3.1/3)$ supersymmetry of the TPM. Recall
that in the case of the TIM, the $\phi _{1,3}$ field scaling as
$3/5+3/5$ conformal object breaks spontaneously the $(1/2,1/2)$
supersymmetry of the $c=7/10$ model. Unfortunately, this feature
cannot be checked directly on the scalar potential $V(\varphi
,\varphi ^{\ast })$ since the theory we are considering in this
section is non unitary .we shall not pursue this direction.
$D=2(1/3.1/3)$ supersymmetry breaking will be discussed on the
following unitary model.\\
\textbf{Acknowledgments}\\ {\small MBS would like to thank the
Abdus Salam International Center for Theoretical Physics (ICTP)for
hospitality during the participation to the workshop on Integrable
systems and scientific computing (15-20 June 2009).


\begin{thebibliography}{99}
\bibitem{1} G.~W.~Semenoff,
%``Canonical Quantum Field Theory with Exotic Statistics,''
Phys.\ Rev.\ Lett.\ \textbf{61}, 517 (1988);\newline
C. Ahn, D. Bernard, A. LeClair, Nucl Phys. B 346 (1990)409;\newline
D. Bernard and A. LeClair, Nucl Phys. B 340 (1990)721;\newline
A.B. Zamolodchikov, Int. J. Mod. Phys. A4 (1989)4235. %\cite{2}

\bibitem{ssz1} E.H.Saidi, M.B.Sedra, J.Zerouaoui,
%``On D = 2 (1/3, 1/3) supersymmetric theories 1,''
Class.\ Quant.\ Grav.\ \textbf{12} (1995) 1567;

\bibitem{ssz2} E.H.Saidi, M.B.Sedra, J.Zerouaoui,
%``On D = 2 (1/3, 1/3) supersymmetric theories 2,''
Class.\ Quant.\ Grav.\ \textbf{12}(1995) 2705.
%%CITATION = CQGRD,12,2705;%%

\bibitem{4} A.~Perez, M.~Rausch de Traubenberg, P.~Simon,
%``$2D$ Fractional Supersymmetry for Rational Conformal Field Theory.
%Application for Third-Integer Spin States,''
Nucl.Phys.B 482(1996)325; %%CITATION = NUPHA,B482,325;%%
\bibitem{5}M.~Rausch de Traubenberg, P.~Simon,
%``2-D fractional supersymmetry and conformal field theory for alternative
%statistics,''
Nucl.Phys.B 517(1998)485. %%CITATION = NUPHA,B517,485;%%
%\cite{sz1}
\bibitem{sz1} M.B.Sedra, J.Zerouaoui,
%``F-Susy And The Three States Potts Model,'
Adv. Studies Theor. Phys., Vol. 2, 2008, N.20, 965 - 973
%\cite{sz2}
\bibitem{sz2} M.B.Sedra, J.Zerouaoui,
%``Heterotic  Fractional Susy Models,'
Adv. Studies Theor. Phys., Vol. 3, 2009, no. 7, 273 - 281
\end{thebibliography}
\end{document}